\documentclass[11pt,twoside]{article}
\pdfoutput=1

\usepackage{amsfonts}
\usepackage{amsmath}
\usepackage{amsthm}
\usepackage{amssymb}
\usepackage{amscd}
\usepackage{eucal}
\usepackage{bm}
\usepackage{graphicx}

\usepackage{hyperref}
\usepackage{a4}
\usepackage{times}
\usepackage{cite}
\usepackage{microtype}

\def\beq{\begin{equation}}
\def\eeq{\end{equation}}
\def\bea{\begin{eqnarray}}
\def\eea{\end{eqnarray}}

\def\beann{\begin{eqnarray*}}
\def\eeann{\end{eqnarray*}}

\let\a=\alpha \let\be=\beta \let\g=\gamma \let\de=\delta
\let\e=\varepsilon  \let\h=\eta 

  \let\la=\lambda \let\m=\mu
\let\n=\nu \let\x=\xi \let\p=\pi \let\r=\rho \let\s=\sigma
\let\om=\omega \let\ps=\psi
   \let\Ps=\Psi

\let\La=\Lambda  \let\D=\Delta

\let\qd=\quad  

\def\epp{\, .}
\def\epc{\, ,}

\def\tst#1{{\textstyle #1}}

\theoremstyle{plain}
\newtheorem{theorem}{Theorem}
\newtheorem*{theorem*}{Theorem}
\newtheorem{lemma}{Lemma}
\newtheorem*{lemma*}{Lemma}

\newtheorem*{proposition*}{Proposition}

\newtheorem*{corollary*}{Corollary}

\newtheorem*{conjecture*}{Conjecture}

\theoremstyle{definition}

\newtheorem*{question*}{Question}

\def\2{\frac{1}{2}} \def\4{\frac{1}{4}}

\def\6{\partial}

\def\+{\dagger}

\def\<{\langle} \def\>{\rangle}

\let\auf=\uparrow \let\ab=\downarrow

\def\i{{\rm i}}

\def\rd{{\rm d}}

\def\ctg{\, {\rm ctg}\,}

\DeclareMathOperator{\re}{e}

\DeclareMathOperator{\sh}{sh}
\DeclareMathOperator{\ch}{ch}

\DeclareMathOperator{\cth}{cth}

\DeclareMathOperator{\arctg}{arctg}

\DeclareMathOperator{\tr}{tr}

\DeclareMathOperator{\sign}{sign}
\DeclareMathOperator{\End}{End}
\DeclareMathOperator{\id}{id}

\renewcommand{\Re}{\operatorname{Re}}
\renewcommand{\Im}{\operatorname{Im}}

\def\av{\mathbf{a}}

\def\ev{\mathbf{e}}

\def\kv{\mathbf{k}}

\def\xv{\mathbf{x}}
\def\yv{\mathbf{y}}

\def\lav{{\boldsymbol{\lambda}}}

\def\fa{\mathfrak{a}}

\allowdisplaybreaks

\begin{document}

\thispagestyle{empty}

\begin{center}

{\LARGE \bf
Bethe Ansatz}

\vspace{18mm}

{\large  \sc
Frank G\"{o}hmann
}\\[3.5ex]
Fakult\"at f\"ur Mathematik und Naturwissenschaften\\
Bergische Universit\"at Wuppertal\\
42097 Wuppertal

\vspace{115mm}

{\large {\bf Abstract}}

\end{center}

\begin{list}{}{\addtolength{\rightmargin}{9mm}
               \addtolength{\topsep}{-5mm}}
\item
The term Bethe Ansatz stands for a multitude of methods in
the theory of integrable models in statistical mechanics
and quantum field theory that were designed to study the
spectra, the thermodynamic properties and the correlation
functions of these models non-perturbatively. This essay
attempts to a give a brief overview of some of these methods
and their development, mostly based on the example of the
Heisenberg model and the corresponding six-vertex model.
\end{list}

\clearpage

\section{Introduction}
The full meaning of the term Bethe Ansatz as it is commonly
used is not easy to grasp. Conceived in a broader sense, it
comprises a large number of methods, developed over the years in
order to analyse model systems rooted in classical and quantum
statistical mechanics or in quantum field theory. In a more narrow
sense the word refers to work of H. Bethe \cite{Bethe31} on the
Heisenberg spin chain in which these methods have their origin.
Starting with Bethe's original work, which is still of great
illustrative value, we shall try to a give a brief overview of
the development of the method. Taking the vast amount of research
literature on this subject, our attempt to stay brief will
necessarily limit us to a few examples, and we will be unable
to do more than just scratching the surface of our subject. For
more in-depth information we have to refer the reader to one
of several readable monographs \cite{KBIBo,Babook,thebook,Gaudin83,%
Takahashi99,Slavnovbook,EckleBa} that have appeared over the years.

\subsubsection*{Basic notations}
For simplicity we shall restrict ourselves in the explicit
examples below mostly to systems with local degrees of freedom in
a finite dimensional Hilbert space ${\cal H} = {\mathbb C}^d$,
equipped with the canonical Hermitian scalar product. We fix
a basis
\begin{equation}
     \{e_\a\}_{\a = 1}^d \subset {\mathbb C}^d \epp
\end{equation}
Then the set $\{e_\a^\be\}_{\a, \be = 1}^d \subset
\End {\mathbb C}^d$, defined by
\begin{equation}
     e_\a^\be e_\g = \de_\g^\be e_\a
\end{equation}
for $\a, \be, \g = 1, \dots, d$ (or $\a, \be, \g = -, +$, if $d = 2$),
is a basis of $\End {\mathbb C}^d$. In this basis the identity
$I_d \in \End {\mathbb C}^d$ has the expansion $I_d = e_\a^\a$. Here
and in the following summation over double Greek indices is implied.

The space of states of the quantum spin chains considered below is the
tensor product space ${\cal H}_L = ({\mathbb C}^d)^{\otimes L}$.
The number of factors $L$ is called the number of lattice
sites or the length of the quantum chain. The embedding of
the basis of elementary endomorphisms
$\{e_\a^\be\}_{\a, \be = 1}^d$ into $\End
({\mathbb C}^d)^{\otimes L}$ is defined by
\begin{equation}
     {e_j}_\a^\be = I_d^{\otimes (j - 1)} \otimes e_\a^\be
                    \otimes I_d^{\otimes (L - j)} \epc
\end{equation}
where $j = 1, \dots, L$. This definition allows us to introduce 
`$m$-site operators'. For every $A \in \End ({\mathbb C}^d)^{\otimes m}$,
$m \le L$, and $\{j_1, \dots, j_m\} \subset \{1, \dots, L\}$ we set
\begin{equation}
     A_{j_1, \dots, j_m} = A^{\a_1 \dots \a_m}_{\be_1 \dots \be_m}
                         {e_{j_1}}_{\a_1}^{\be_1} \dots {e_{j_m}}_{\a_m}^{\be_m} \epp
\end{equation}
We say that $A$ acts non-trivially only on sites $j_1, \dots, j_m$.

Examples of single-site operators for $d = 2$ are the spin
operators
\begin{equation}
     s_j^\a = \2 \s_j^\a \epc \qd \a = x, y, z \epc
\end{equation}
where the $\s^\a$ are the Pauli matrices. Important examples
of two-site operators for arbitrary $ d > 1$ are the
transposition operators $P_{j, k}$, where $P \in
\End \bigl({\mathbb C}^d \otimes {\mathbb C}^d\bigr)$
is defined as
\begin{equation} \label{transmat}
     P = e_\a^\be \otimes e_\be^\a \epp
\end{equation}
For every $\xv, \yv \in {\mathbb C}^d$ we have $P \, \xv \otimes \yv
= \yv \otimes \xv$. The Operators $P_{j, j + 1}$, $j = 1, \dots, L - 1$,
generate a representation of the symmetric group $\mathfrak{S}^L$
acting on ${\cal H}_L$ and (adjointly) on $\End {\cal H}_L$.

\subsubsection*{The Heisenberg Hamiltonian and its symmetries}
The Hamiltonian $H_L \in \End {\cal H}_L$ of the $GL(d)$
Heisenberg model is defined as
\begin{equation} \label{hxxxgen}
     H_L = \frac{J}{2} \sum_{j = 1}^L \bigl(P_{j, j + 1} - 1\bigr) \epc
\end{equation}
where $J \in {\mathbb R}$ and where periodic boundary conditions,
$P_{L, L + 1} = P_{L, 1}$, are implied . This model can be
`solved' (in a way explained below) by Bethe Ansatz for any
$d > 1$. If $d = 2$ the Hamiltonian (\ref{hxxxgen}) is called
the Heisenberg or XXX model. This is the model originally treated
by Bethe in \cite{Bethe31}. If $d = 2$ the transposition operator
$P$ can be neatly expressed in terms of spin operators, and $H_L$
takes its most familiar form
\begin{equation} \label{hxxx}
     H_L = J \sum_{j = 1}^L \bigl(s_j^\a s_{j+1}^\a - \tst{\4}\bigr) \epp
\end{equation}
The Heisenberg model, generally defined on any crystal lattice
of dimension 1, 2 or 3, is the fundamental model for the
antiferromagnetism of matter. It can be obtained by applying
second order degenerate perturbation theory at strong coupling
to the underlying Hubbard model (see e.g.\ \cite{Goehmann21},
lectures 20, 21). Considered on a one-dimensional periodic
lattice it is one of the rare examples of an interacting many-body
quantum system, that can be understood to a large extent without
recourse to any further simplifying restriction or approximation.

The spatial symmetry group of $H_L$ is the dihedral group
${\cal D}_L$ which is the symmetry group of a regular polyhedron
with $L$ edges. It is generated as a semidirect product by the
two cyclic elements
\begin{subequations}
\begin{align} \label{defu}
     \hat U & = P_{1,2} \dots P_{L-1, L} \epc \\[1ex]
     \hat P & = P_{L/2, L/2 + 1} \dots P_{1,L} \epc
\end{align}
\end{subequations}
the shift operator and the parity operator. Here we have assumed
for simplicity that $L$ is even. Except for its spatial symmetry
$H_L$ has an `external' $\mathfrak{sl}_2$ symmetry. If we define
the total spin operators
\begin{equation}
     S^\a = \sum_{j=1}^L s_j^\a \epc
\end{equation}
Then
\begin{equation}
     [S^\a,S^\be] = \i \e^{\a \be \g} S^\g \epc \qd
        [S^\a, \hat U] = [S^\a, H_L] = 0
\end{equation}
for $\a, \be = x, y, z$. We may therefore construct simultaneous
eigenvectors of the operators, $H_L$, $\hat U$, $S^z$ and $(S^\a)^2$.
This is what the Bethe Ansatz does.
\section{Quantum chains}
Quantum chains are one-dimensional lattice models of quantum
mechanics. A sub-class are the quantum spin chains for which
the local degrees of freedom at every lattice site are
quantum spins. An example is the Heisenberg model with Hamiltonian
(\ref{hxxx}).
\subsection{Bethe's work and some of its consequences}
\subsubsection*{\boldmath $S^z$ eigenbasis and wavefunctions}
The first step in the Bethe Ansatz analysis of the Heisenberg model
is to utilize the $S^z$ invariance of the Hamiltonian (\ref{hxxx}) by
switching to an appropriate basis. This basis is constructed as as follows.
We define the ferromagnetic or pseudo vacuum state
\begin{equation} \label{pseudo}
     |0\> = e_+^{\otimes L} \epp
\end{equation}
For any $\xv = \sum_{j=1}^N x_j \ev_j$ with canonical unit row
vectors $\ev_j$ and $1 \le x_1 < \dots < x_N \le L$ we define
\begin{equation}
     |\xv\> = s_{x_N}^- \dots s_{x_1}^- |0\> \epc
\end{equation}
where $s^- = s^x - \i s^y$ is the spin lowering operator. Clearly
there are $\sum_{N=0}^L \binom{L}{N} = 2^L$ such states. These
are linearly independent, hence form a basis ${\cal B}_z \subset
{\cal H}_L$. For any $|\xv\> = |(x_1, \dots, x_N)\> \in {\cal B}_z$
\begin{equation}
     S^z|\xv\> = \tst{\2}(L - 2N)|\xv\> \epp
\end{equation}
Thus, ${\cal B}_z$ is a basis of $S^z$ eigenstates, and $H_L$
is block diagonal in this basis.

For this reason we can diagonalize $H_L$ for a fixed eigenvalue
$L/2 - N$ of $S^z$ or `in a sector with a fixed number $N$ of
overturned spins'. Let $|\Ps\>$ be any state with $S^z |\Ps\> =
(L/2 - N) |\Ps\>$. We shall call the coefficients $\Ps (\xv)$
in the expansion
\begin{equation}
     |\Ps\> = \sum_{1 \le x_1 < \dots < x_N \le L} \Ps(\xv) |\xv\>
\end{equation}
the ($N$-particle) wave functions. Note that in this notation
$\<\xv|\Ps\> = \Ps(\xv)$.

\subsubsection*{Eigenvalue problem for the wavefunctions}
The next step is to recognize the following simple lemma that
follows directly from the action of the Hamiltonian on the
basis ${\cal B}_z$.
\begin{lemma} \label{lem:firstqh}
The eigenvalue problem
\begin{equation} \label{ewp}
     H_L |\Ps\> = E |\Ps\>
\end{equation}
for the Heisenberg Hamiltonian (\ref{hxxx}) in the sector
of $N$ overturned spins is equivalent to the following set
of conditions on the corresponding $N$-particle wave functions
$\Ps(\xv)$.
\begin{enumerate}
\item
Wave equation
\begin{equation} \label{waveeqn}
     \frac{J}{2} \sum_{j=1}^N
        \bigl(\Ps (\xv + \ev_j) - 2 \Ps(\xv)
	      + \Ps(\xv - \ev_j)\bigr) = E \Ps(\xv) \epp
\end{equation}
\item
Reflection condition
\begin{equation} \label{reflect}
     \Ps(\xv + \ev_j) - 2 \Ps(\xv) + \Ps(\xv - \ev_{j+1}) = 0
\end{equation}
for every $j \in \{1, \dots, N\}$ for which $x_j + 1 = x_{j+1}$.
\item
Periodic boundary conditions for $x_1 = 0$ and $x_N = L + 1$,
\begin{subequations}
\label{period}
\begin{align} \label{perioda}
     \Ps(0, x_2, \dots, x_N) & = \Ps(x_2, \dots, x_N, L) \epc \\[1ex] \label{periodb}
     \Ps(x_1, \dots, x_{N-1}, L+1) & = \Ps(1, x_1, \dots, x_{N-1}) \epp
\end{align}
\end{subequations}
\end{enumerate}
\end{lemma}
A good way of thinking of these equations is that for a given
down-spin configuration $\xv$ some of the shifted configurations
$\xv \pm \ev_j$ in (\ref{waveeqn}) may be located outside the
`simplex' $1 \le x_1 < \dots < x_N \le L$ and that those wrong
configurations are reduced to the correct ones by (\ref{reflect}),
(\ref{period}). Such `exceptional configurations' occur in
(\ref{waveeqn}) precisely if $x_1 = 1$, $x_N = L$ or if two
neighbouring down-spin positions $x_j$, $x_{j+1}$ differ by $1$.

\subsubsection*{Solving the eigenvalue problem}
In the following we shall make frequent use of the natural right
action of the symmetric group $\mathfrak{S}^N$ on row vectors
with entries in ${\mathbb C}$. For any such $\kv = (k_1, \dots, k_N)$
we define
\begin{equation}
     \kv Q = (k_{Q1}, \dots, k_{QN}) \qd \text{for all $Q \in \mathfrak{S}^N$.}
\end{equation}
This defines a representation of $\mathfrak{S}^N$ which is
orthogonal with respect to the Euclidian scalar product
$\<\kv, \xv\> = \sum_{j=1}^N k_j x_j$. Namely, $\<\kv, \xv Q\>
= \<\kv Q^{-1}, \xv\>$.

The next step is now to find solutions of the wave equation
(\ref{waveeqn}) that satisfy the reflection condition (\ref{reflect}).
Every function $\re^{\i \<\kv, \xv\>}$ for $\kv \in {\mathbb C}^N$
arbitrary satisfies (\ref{waveeqn}) and is degenerate with
$\re^{\i \<\kv Q, \xv\>}$ for every $Q \in \mathfrak{S}^N$.
It follows that
\begin{equation} \label{ba}
     \Ps (\xv) = \sum_{Q \in \mathfrak{S}^N} A(Q) \re^{\i \<\kv Q, \xv\>} \epc
\end{equation}
where the $A(Q) \in {\mathbb C}$ are arbitrary amplitudes, gives a
solution of the wave equation with
\begin{equation}
     E = E(\kv) = J \sum_{j=1}^N \bigl(\cos(k_j) - 1\bigr) \epp
\end{equation}
At the same time $\Ps (\xv)$ is a solution of the eigenvalue
equation of the shift operator $\hat U \Ps (\xv) = \om \Ps (\xv)$
with eigenvalue
\begin{equation}
     \om = \om(\kv) = \re^{\i \sum_{j=1}^N k_j} \epp
\end{equation}

The wave function (\ref{ba}) is Bethe's Ansatz\footnote{Ansatz is
a German word for trial function or substitution (e.g.\ for a
differential equation or an eigenvalue problem) typically depending
on a certain number of parameters. Substituting the trial function
into the equations under consideration implies a set of necessary
conditions for the parameters.} or `the Bethe Ansatz wave function'.
Note that the Bethe Ansatz wave function is not the general solution
of the wave equation for fixed $E$. In general we might superpose
waves with $\kv' \ne \kv$ but $E(\kv') = E(\kv)$. A further restriction
on $\kv$ then comes from the fact that $\hat U$ is conserved,
requiring that $\om(\kv') = \om(\kv)$. We shall see below that
there exist more conserved quantities with operators $\hat I$ 
commuting with $H_L$ which have a spectrum of the same type. This
imposes more conditions of the type $I(\kv') = I(\kv)$ on the
`quasi momenta' $\kv$, and we may imagine that sufficiently
many constraints finally imply that $\kv' = \kv$. Still, we are
not aware of any rigorous argument of this type.

In any case, as we shall see, the Bethe Ansatz wave function permits
us to satisfy the reflection condition (\ref{reflect}). For any
$j = 1, \dots, N-1$ let $\Pi_j \in \mathfrak{S}^N$ be the transposition
of $j$ and $j+1$. Then $x_j + 1 = x_{j+1}$ implies $(\xv + \ev_j)\Pi_j
= \xv + \ev_j$, $(\xv - \ev_{j+1})\Pi_j = \xv - \ev_{j+1}$, and, inserting
(\ref{ba}) into (\ref{reflect}), we obtain
\begin{align}
     \sum_{Q \in \mathfrak{S}^N} & A(Q)
        \Bigl(\re^{\i \<\kv Q, \xv + \ev_j\>} - 2 \re^{\i \<\kv Q, \xv\>}
	         + \re^{\i \<\kv Q, \xv - \ev_{j+1}\>}\Bigr) \notag \\
     & = \sum_{Q \in \mathfrak{A}^N} \Bigl\{
            A(Q) \Bigl(\re^{\i \<\kv Q, \xv + \ev_j\>} - 2 \re^{\i \<\kv Q, \xv\>}
	                  + \re^{\i \<\kv Q, \xv - \ev_{j+1}\>}\Bigr) \notag \\[-1ex]
     & \mspace{72.mu} +
            A(Q\Pi_j) \Bigl(\re^{\i \<\kv Q\Pi_j, \xv + \ev_j\>} - 2 \re^{\i \<\kv Q\Pi_j, \xv\>}
	                  + \re^{\i \<\kv Q\Pi_j, \xv - \ev_{j+1}\>}\Bigr)\Bigr\} \notag \\[1ex]
     & = \sum_{Q \in \mathfrak{A}^N} \Bigl\{
            A(Q) \Bigl(\re^{\i k_{Qj}} - 2 + \re^{- \i k_{Q(j+1)}}\Bigr) \notag \\[-1ex]
     & \mspace{72.mu} +
            A(Q\Pi_j) \Bigl(\re^{\i k_{Qj}} - 2 \re^{\i \<\kv Q, \xv (\Pi_j - \id)\>}
	                  + \re^{- \i k_{Q(j+1)}}\Bigr)\Bigr\} \re^{\i \<\kv Q, \xv\>} = 0 \epp
\end{align}
Here we used in the first equation that $\mathfrak{S}^N$ has
the coset decomposition $\mathfrak{S}^N = \mathfrak{A}^N \cup
\mathfrak{A}^N \Pi_j$, where $\mathfrak{A}^N$ is the alternating
group of order $N$. In the second equation we used that the
representation of the symmetric group on row vectors is
orthogonal with respect to the Euclidian scalar product and
that $\Pi_j^{-1} = \Pi_j$. Finally, using that $\<\kv Q, \xv (\Pi_j - \id)\>
= k_{Qj} - k_{Q(j+1)}$, if $x_j + 1 = x_{j+1}$, we see that
the reflection condition is satisfied, if
\begin{equation}
     \frac{A(Q\Pi_j)}{A(Q)} =
        - \frac{\re^{\i(k_{Qj} + k_{Q(j+1)})} - 2 \re^{\i k_{Q(j+1)}} + 1}
               {\re^{\i(k_{Qj} + k_{Q(j+1)})} - 2 \re^{\i k_{Qj}} + 1} \epp
\end{equation}

The right hand side of this equation simplifies under a change
of variables from quasi momenta $k$ to so-called rapidities
\cite{Hulthen38}
\begin{equation}
     \la = \2 \ctg\Bigl(\frac{k}{2}\Bigr)\ \Leftrightarrow\
        \re^{\i k} = \frac{\la + \frac{\i}{2}}{\la - \frac{\i}{2}} \epp
\end{equation}
In these variables
\begin{equation} \label{arec}
     \frac{A(Q\Pi_j)}{A(Q)} =
        \frac{\la_{Qj} - \la_{Q(j+1)} - \i}{\la_{Qj} - \la_{Q(j+1)} + \i} =
        \frac{\la_{Q\Pi_j (j + 1)} - \la_{Q\Pi_j j} - \i}{\la_{Q\Pi_j (j+1)} - \la_{Q\Pi_j j}}
        \frac{\la_{Q (j+1)} - \la_{Qj}}{\la_{Q (j + 1)} - \la_{Qj} - \i} \epp
\end{equation}
Now
\begin{equation}
        \frac{\la_{Q (j+1)} - \la_{Qj}}{\la_{Q (j + 1)} - \la_{Qj} - \i}
	   \prod_{1 \le \ell < k \le N}
	   \frac{\la_{Q\ell} - \la_{Qk} - \i}{\la_{Q\ell} - \la_{Qk}}
\end{equation}
is invariant under the replacement $Q \leftrightharpoons Q \Pi_j$
by inspection, implying that
\begin{equation} \label{aexpl}
     A(Q) = \prod_{1 \le \ell < k \le N}
            \frac{\la_{Q\ell} - \la_{Qk} - \i}{\la_{Q\ell} - \la_{Qk}}
\end{equation}
is a solution of (\ref{arec}).

Thus, with $A(Q)$ according to (\ref{aexpl}) the Bethe
Ansatz wave function (\ref{ba}) satisfies the wave equation
(\ref{waveeqn}) and the reflection condition (\ref{reflect}),
still for an arbitrary set of rapidities $\{\la_1, \dots,
\la_N\} \subset {\mathbb C}$. The last step is now to use
the periodic boundary condition (\ref{period}) in order to
impose restrictions on these sets of variables. It turns out
that (\ref{perioda}) and (\ref{periodb}) lead to the same
equations. For this reason we focus on (\ref{perioda}).
Inserting the Bethe Ansatz wave function we obtain
\begin{multline} \label{bapbc}
     0 = \sum_{Q \in \mathfrak{S}^N} A(Q)
        \bigl\{\re^{\i \<\kv Q, \xv\>}
	       - \re^{\i \<\kv Q, \xv U_N + L \ev_N\>}\bigr\}\bigr|_{x_1 = 0} \\
     = \sum_{Q \in \mathfrak{S}^N}
        \bigl\{A(Q) - A(Q U_N) \re^{\i k_{Q1} L}\bigr\}
	   \re^{\i \<\kv Q, \xv\>} \bigr|_{x_1 = 0} \epc
\end{multline}
where $U_N$ is the order-$N$ cyclic element, $U_N j = (j + 1) \mod N$.
Eq.~(\ref{bapbc}) is obviously satisfied, if
\begin{equation} \label{cycle}
     \frac{A(Q U_N)}{A(Q)} = \re^{- i k_{Q1} L} \qd \text{for all $Q \in \mathfrak{S}^N$.}
\end{equation}

The cyclic element $U_N$ has the decomposition $U_N = \Pi_1
\Pi_2 \dots \Pi_{N-1}$. Using (\ref{arec}) we can therefore
conclude that
\begin{multline}
     \frac{A(Q U_N)}{A(Q)} =
        \frac{A(Q \Pi_1)}{A(Q)}
	\frac{A(Q \Pi_1 \Pi_2)}{A(Q \Pi_1)} \dots
	\frac{A(Q \Pi_1 \dots \Pi_{N-1})}{A(Q \Pi_1 \dots \Pi_{N-2})} \\
	= \frac{\la_{Q1} - \la_{Q2} - \i}{\la_{Q1} - \la_{Q2} + \i}
	  \frac{\la_{Q1} - \la_{Q3} - \i}{\la_{Q1} - \la_{Q3} + \i} \dots
	  \frac{\la_{Q1} - \la_{QN} - \i}{\la_{Q1} - \la_{QN} + \i}
	= - \prod_{\ell=1}^N \frac{\la_{Q1} - \la_\ell - \i}{\la_{Q1} - \la_\ell + \i} \\
        = \re^{- i k_{Q1} L}
	= \biggl(\frac{\la_{Q1} - \frac{\i}{2}}{\la_{Q1} + \frac{\i}{2}}\biggr)^L
\end{multline}
for all $Q \in \mathfrak{S}^N$. The latter is equivalent to
\begin{equation} \label{bae}
	\biggl(\frac{\la_j - \frac{\i}{2}}{\la_j + \frac{\i}{2}}\biggr)^L
	= - \prod_{k=1}^N \frac{\la_j - \la_k - \i}{\la_j - \la_k + \i} \qd
	\text{for $j = 1, \dots, N$.}
\end{equation}
These are the so-called Bethe Ansatz equations \cite{Bethe31}.
They determine the allowed sets $\{\la_1, \dots, \la_N\}$ of
rapidities for which the Bethe Ansatz wave function (\ref{ba})
with amplitudes (\ref{aexpl}) provide solutions of the
eigenvalue problem (\ref{ewp}).

We have thus shown the following
\begin{theorem} \label{thm:cba}
Bethe 1931 \cite{Bethe31}. The Heisenberg Hamiltonian (\ref{hxxx})
has a set of eigenstates
\begin{multline} \label{baev}
     |\{\la_j\}\> = \\[-1ex] \sum_{1 \le x_1 < \dots < x_N \le L}
        \sum_{Q \in \mathfrak{S}^N}
	\biggl[\prod_{1 \le k < \ell \le N}
	   \frac{\la_{Qk} - \la_{Q \ell} + \i}{\la_{Qk} - \la_{Q \ell}}\biggr]
        \biggl[\prod_{k=1}^N
	   \biggl(\frac{\la_{Qk} + \frac{\i}{2}}{\la_{Qk} - \frac{\i}{2}}\biggr)^{x_k}
	   s_{x_k}^-\biggr] |0\>
\end{multline}
with eigenvalues
\begin{equation} \label{eev}
     E = - \frac{J}{2} \sum_{j=1}^N \frac{1}{\la_j^2 + \frac{1}{4}} \epc
\end{equation}
where the $\la_j$, $j = 1, \dots, N$, are solutions of the Bethe
Ansatz equations (\ref{bae}). The corresponding lattice momenta are
\begin{equation} \label{pev}
     P = \biggl[- \i \sum_{j=1}^N
                  \ln\biggl(\frac{\la_j + \frac{\i}{2}}{\la_j - \frac{\i}{2}}\biggr)\biggr]
		  \mod 2\p \epc
\end{equation}
the corresponding eigenvalues of $S^z$ are $L/2 - N$.
\end{theorem}

\subsubsection*{Highest-weight property}
Much can be said about the interpretation and the scope of this
result.  First of all the Bethe eigenstates (\ref{baev}) are 
highest-weight vectors of irreducible $\mathfrak{sl}_2$ representations.
For $N = 0, \dots, L$ they satisfy
\begin{equation} \label{slzwohw}
     S^+ |\Ps\> = 0 \epc
\end{equation}
where $S^+ = S^x + \i S^y$ is the raising operator of the total
spin. This can be seen \cite{Gaudin83} by using the cyclicity
condition (\ref{cycle}) which is equivalent to the Bethe Ansatz
equations (\ref{bae}). If $S^- = S^x - \i S^y$ is the total
spin-lowering operator, we conclude that
\begin{equation}
     0 \le \<\Ps|\bigl(S^x S^x + S^y S^y\bigr)|\Ps\> =
           \<\Ps|\bigl(S^- S^+ + S^z\bigr)|\Ps\> = (L/2 - N) \|\Ps\|^2 \epp
\end{equation}
Hence $|\Ps\> = 0$ for $N > L/2$ if the Bethe Ansatz equations
(\ref{bae}) are satisfied, while in this case $|\Ps\>$ for
$N \le L/2$ is associated with a whole $\mathfrak{sl}_2$ multiplet
$(S^-)^j |\Ps\>$, $j = 0, \dots, L - 2N$, of degenerate eigenstates.

\subsubsection*{Admissible solutions, off-shell Bethe vectors and
the completeness problem}
The next observation is that,
with our choice (\ref{aexpl}), the amplitude is not defined if
two of the rapidities $\la_j$, $\la_k$, $j \ne k$,  coincide or
if $\la_j = \frac{\i}{2}$ for some $j \in \{1, \dots, N\}$.
This deficiency may be cured by multiplying all amplitudes $A(Q)$
by the Vandermonde determinant $\prod_{1 \le k < \ell \le N}
(\la_k - \la_\ell)$ and by $\prod_{j=1}^N (\la_j - \frac{\i}{2})^{L+1}$.
The resulting Bethe wave function in the new normalization is
\begin{multline} \label{wavefunren}
     \Ps \bigl(\xv\big|\{\la_j\}_{j=1}^N\bigr) = \\
        \sum_{Q \in \mathfrak{S}^N} \sign(Q)
	\biggl[\prod_{1 \le k < \ell \le N}
	   (\la_{Qk} - \la_{Q \ell} + \i)\biggr]
        \prod_{k=1}^N
	   \bigl(\la_{Qk} + \tst{\frac{\i}{2}}\bigr)^{x_k}
	   \bigl(\la_{Qk} - \tst{\frac{\i}{2}}\bigr)^{L - x_k + 1} \epp
\end{multline}
This wave function is now regular if two rapidities coincide
or if a rapidity equals $\frac{\i}{2}$. However, if e.g.\
$\la_{k_1} = \la_{k_2}$ for $k_1 \ne k_2$, the two products
on the right hand side of (\ref{wavefunren}) are invariant
under the substitution $Q \leftrightharpoons \Pi_{k_1, k_2} Q$,
where $\Pi_{k_1, k_2}$ is the transposition of $k_1$ and $k_2$,
and the wave function vanishes. It also vanishes if $\la_j - \la_k
= \pm \i$ or if $\la_j = \pm \frac{\i}{2}$. The Bethe
wave function $\Ps \bigl(\xv\big|\{\la_j\}_{j=1}^N\bigr)$
satisfies the wave equation (\ref{waveeqn}) and the reflection
condition (\ref{reflect}) for arbitrary sets $\{\la_j\}_{j=1}^N
\subset {\mathbb C}$. In this unrestricted case it is called an
`off-shell' Bethe wave function, and we can build the corresponding
off-shell Bethe vector
\begin{equation} \label{bvoffshell}
     |\{\la_j\}\> = \sum_{1 \le x_1 < \dots < x_N \le L}
        \Ps \bigl(\xv\big|\{\la_j\}_{j=1}^N\bigr) |\xv\> \epp
\end{equation}
Solutions of the Bethe Ansatz equations (\ref{bae}) for
which all rapidities are mutually distinct and for which
$\la_j - \la_k \ne \i$ and $\la_j \ne \pm \frac{\i}{2}$ for
all $j,k = 1, \dots, N$ are called admissible solutions.
Thus, $|\{\la_j\}\>$ can only be an eigenvector of the Hamiltonian
$H_L$, if $\{\la_j\}_{j=1}^N$ is an admissible solution of
the Bethe Ansatz equations (\ref{bae}) and if $N \le L/2$.
Such eigenvectors are then $\mathfrak{sl}_2$ highest-weight
vectors. A natural question is, if the corresponding multiplets
associated with all admissible solutions of the Bethe Ansatz
equations (\ref{bae}) generate a basis of ${\cal H}_L$.
Interestingly, the answer is negative as can be seen
by considering one of the simplest examples, $L = 4$.

\subsubsection*{Completeness}
In order to obtain an eigenbasis it is sufficient to consider
limits of normalized Bethe wave functions. There are two
possibilities.  One may take off-shell Bethe wave functions and
send the rapidities to an inadmissible solution of the Bethe
equations or to infinity. A problem in this case is to decide
which inadmissible solutions are appropriate. This problem
was solved in \cite{MTV09} by classifying solutions of
Baxter's $TQ$ equation, rather than solutions of the Bethe
Ansatz equations. They are under control by means of
the representation theory of the Yangian quantum group
$Y(\mathfrak{gl}_2)$ and can be counted and mapped bijectively
to eigenvectors of $H_L$. Another possible way to construct
the missing eigenvectors is by introducing an on-shell
regularization, which makes all solutions of the Bethe Ansatz
equations admissible, and then sending the regularization 
parameters to zero. The regularization must be such that
the $\mathfrak{sl}_2$ invariance is preserved and such that
the regularized model can still be solved by a similar Bethe
Ansatz procedure as before. As we shall see below, the Heisenberg
Hamiltonian commutes with the transfer matrix of the rational
six-vertex model. The latter naturally carries $L$ so-called
inhomogeneity parameters, which provide an appropriate
regularization of the Hamiltonian. Nevertheless, even in
this setting, a proof \cite{MTV09} of `the completeness of
the Bethe Ansatz' remains highly non-trivial and involves
many of the modern tools that originated from the Bethe Ansatz,
such as the representation theory of the quantum group
$Y(\mathfrak{gl}_2)$ \cite{Drinfeld85, Drinfeld87b,ChPr90},
Baxter's $TQ$ equation \cite{Babook}, or Sklyanin's method
of the separation of variables \cite{Sklyanin92}. The
generalized Bethe vectors of the Hamiltonian $H_L$ are
finally obtained by performing the homogeneous limit for
the normalized eigenvectors of the inhomogeneous transfer
matrix. In the general case this brings about derivatives
of the Bethe wave functions with respect to the inhomogeneity
parameters.

\subsubsection*{Ferromagnetic ground state and magnons}
Theorem~\ref{thm:cba} is valid for either sign of the real
parameter $J$ which has the physical meaning of an `exchange
interaction'. If $J < 0$ the model is called ferromagnetic,
if $J > 0$ it is called antiferromagnetic. Both cases are
very different, physically as well as mathematically. Reversing
the sign of $J$ means to `invert the spectrum'. The ground
state of the ferromagnet is the highest excited state of
the antiferromagnet and vice versa. For $d = 2$ the operator
$P^- = (I_4 - P)/2$ is the projector onto the spin-singlet
state. Since $(P^-)^2 = P^-$, its spectrum is the set $\{0, 1\}$,
like for every projection operator. If $J < 0$, then, for every
$|\Ps\> \in {\cal H}_L$,
\begin{equation}
     \<\Ps|H_L|\Ps\> = - J \sum_{j=1}^L \<\Ps|P_{j,j+1}^-|\Ps\> \ge 0 \epc
\end{equation}
where we have estimated each term in the sum by its
smallest possible eigenvalue $0$. Thus, in this case, the
pseudo vacuum state $|0\>$ is a ground state of the
Hamiltonian, which is degenerate with the other states
in the multiplet, $(S^-)^j |0\>$, $j = 1, \dots, L$. These
ground states have maximal spin $S^z = L/2$. They lie in
the sector of no overturned spin, $N = 0$, and their energy
and momentum eigenvalues are $E = 0$, $P = 0$. Hence,
(\ref{eev}) and (\ref{pev}) are the energies and momenta
of excited states relative to the ground state. These
equations hold in particular for $N = 1$. In that case
we obtain excitations which have definite energy, momentum
and have spin equal to 1 (since one spin-$\2$ is flipped).
They are naturally interpreted as spin-$1$ particles called
magnons. Because of the additive structure of
Eqs.~(\ref{eev}) and (\ref{pev}), generic excitations are
interpreted as multi-magnon excitations, scattering states
of magnons for which the interaction among the magnons is
encoded in the Bethe Ansatz equations (\ref{bae}).

\subsubsection*{Scattering of magnons and an interpretation of the Bethe Ansatz equations}
This interpretation can be made more precise by looking
at the Bethe Ansatz wave function (\ref{ba}). It may be
interpreted as a superposition of waves in which particle
$j$ carries momentum $k_{Qj}$ and the momenta are
distributed in all possible ways, labeled by permutations
$Q \in \mathfrak{S}^N$, over the particles. Two particles
that scatter interchange their momenta. The scattering is
such that the full set of (quasi) momenta is conserved in
the scattering. Conservation of the magnitude of the
individual momenta is characteristic of two-particle
scattering and is typically violated if more than two
particles are involved. As mentioned above the conservation
can be attributed to the existence of `higher conserved quantities'.
One says that the multi-particle scattering factorizes into
two-particle scattering processes. This can be further
detailed upon introducing an $S$-matrix. Then the factorization
of the scattering processes translates into the factorization
of the multi-particle $S$-matrix into two-particle $S$-matrices
\cite{Yang68}. In our case at hand, like in potential
scattering in one dimension, the ratio of two amplitudes related
by the interchange of two particles, i.e., the ratio on the left 
hand side of (\ref{arec}), is interpreted as the two-particle
scattering phase. Hence, the factors in the Bethe Ansatz
equations (\ref{bae}) are the two-particle scattering
phases of particle $j$ on any other particle $k \ne j$.
This provides us with an interpretation of the Bethe Ansatz
equations. If $N = 1$ they are of the form of the
quantization conditions of a free particle in a finite
box with periodic boundary conditions. Thus, the interpretation
for $N > 1$ is that these equations are the finite volume
quantization conditions as modified by $N - 1$ two-particle
scattering phases, when particle $j$ is taken once around
the periodic box, thereby scattering on all other particles.

\subsubsection*{Bound states of magnons and the string hypothesis}
The generic low-energy excitations over the ferromagnetic ground
state are excitations of a small finite number of magnons.
It is an instructive exercise \cite{Bethe31,EKS92c} to consider the
sector $N = 2$ of the Bethe Ansatz equations (\ref{bae}).
Unlike for larger $N$, this can still be done by hand. One
of the interesting findings is that the solutions are not all
real. There exist solutions, consisting of two non-real complex
conjugate rapidities. The corresponding wave functions decay
exponentially as a function of the relative coordinate
$x_1 - x_2$ and can therefore be interpreted as two-magnon bound
states. For larger $N$ one may rely, e.g., on numerical analysis
\cite{HaCa07} to recognize that solutions containing up to $N$
non-real rapidities exist and are grouped in complex conjugate
pairs. The invariance of all sets of solutions $\{\la_j\}_{j=1}^N$
under complex conjugation was proved in \cite{Vladimirov86}.
A more precise description of the eigenstates in the
ferromagnetic case is thus, that they are scattering states
of magnons and bound states of magnons.

The rapidities solving the Bethe Ansatz equations (\ref{bae})
are called the Bethe roots. The occurrence of complex
Bethe roots makes the analysis of the solutions of the
Bethe Ansatz equations difficult. In fact, after many
years of research in mathematical physics, no complete
classification scheme of all solutions of the Bethe Ansatz
(\ref{bae}) equations is known. This is closely connected
with the fact that there is no simple proof of completeness.
The same difficulty triggered, in a very fruitful way, many
attempts to get rid of the Bethe Ansatz equations at all,
e.g., by working directly in the infinite volume, some
of which will be discussed below.

Most attempts by physicists to classify the solutions of the
Bethe Ansatz equations centered about the following 
observation. Fix $\e > 0$, $M > 0$ and fix $N$ in (\ref{bae}).
Consider some $\la_j$ with $\Re \la_j < M$ and $\Im \la_j > \e$.
If then $L$ is very large in (\ref{bae}), the left 
hand side is exponentially small in $L$. This can only be
compensated by the right hand side, if for some $k \in
\{1, \dots, N\}$, $k \ne j$, $\la_j - \la_k$ is exponentially
close to $\i$. Taking into account that the Bethe roots come
in complex conjugate pairs, a solution $\{\la_j\}_{j=1}^N$ may
thus contain a subset of Bethe roots of the form $\{\la_\a^n
+ (n + 1 - 2j) \frac{\i}{2} + \de_\a^{n, j} |j = 1, \dots, n\}$,
where the $\de_\a^{n, j}$ are exponentially small in $L$ and
$\la_\a^n \in {\mathbb R}$. Such configurations are called (ideal)
$n$-strings. They were introduced in Bethe's original work
\cite{Bethe31}. Bethe (incorrectly) suggested that they might
provide a complete classification of all solutions of the Bethe
Ansatz equations and based a counting argument on this hypothesis
that implied completeness. However, if $M$ or $N$ grows with $L$
the parameters $\de_\a^{n, j}$ of the string deviations are not
necessarily small, the ideal strings get severely deformed,
and it is unknown how many strings involving $n$ roots exist
for general given $L$ and $N$. We will briefly come back to the
issue of strings, when we discuss the Bethe Ansatz approaches to
the thermodynamics of the Heisenberg chain below.

\subsection{Early developments and extensions}
\subsubsection*{Identification of the antiferromagnetic ground state}
As opposed to the trivial ferromagnetic ground state which
is $(L+1)$-fold degenerate, which has an explicit description and
an energy per lattice site of $e = 0$, the antiferromagnetic
ground state is a highly non-trivial and `strongly correlated'.
Following Lieb, Schultz and Mattis \cite{LSM61} we note that
$H_L$ is unitarily equivalent to
\begin{equation}
     H_L' = J \sum_{j=1}^L \bigl(s_j^z s_{j+1}^z - s_j^x s_{j+1}^x
                                 - s_j^y s_{j+1}^y - \tst{\4}\bigr)
\end{equation}
by one of the transformations $S_e = \prod_{j=1}^{L/2}
\s_{2j}^z$ or $S_o = \prod_{j=1}^{L/2} \s_{2j-1}^z$,
if $L$ is even. Clearly $[H_L',S^z] = 0$; and $L$ even
also implies that $0$ is an eigenvalue of $S^z$. Then
it is not hard to see that $H_L'$ restricted to its
$S^z = 0$ subspace ${\cal H}_{L, 0}$ satisfies the
requirements of the Perron-Frobenius theorem \cite{Meyer00}.
The latter implies that $H_L'$ has a unique translation-invariant
ground state $|g'\>$ on ${\cal H}_{L, 0}$. It follows
that $|g\> = S_e |g'\>$ is the unique ground state of
$H_L$ on ${\cal H}_{L, 0}$ and that
\begin{equation}
     \hat U |g\> = S_o |g'\> = S_e S_e S_o |g'\> = (-1)^{L/2} |g\> \epp
\end{equation}
Thus, $|g\>$ has momentum $P = 0$, if $L$ is divisible
by $4$, and momentum $P = \p$ otherwise. By the Lieb-Mattis
theorem \cite{LiMa62} the ground state of $H_L$ in
${\cal H}_L$ must be a singlet state, so it must be
in ${\cal H}_{L, 0}$. Hence $|g\>$ is the unique ground
state of $H_L$ on ${\cal H}_L$. Due to the completeness of
the Bethe Ansatz $|g\>$ is a (generalized) Bethe vector.

\subsubsection*{Simple observables of the antiferromagnet in the thermodynamic limit}
Taking the logarithm of the Bethe Ansatz equations (\ref{bae}) 
we obtain
\begin{equation} \label{logbae}
     \frac{1}{\p} \arctg (2 \la_j) = \frac{n_j}{L} - \frac{N+1}{2L}
        + \sum_{k=1}^N \frac{1}{\p L} \arctg (\la_j - \la_k) \epc
\end{equation}
where $n_j \in {\mathbb Z}$. Here different sets of
solutions $\{\la_j\}_{j=1}^N$ are parameterized by
different sets of integers $\{n_j\}_{j=1}^N$. The
ground state corresponds to $n_j = j$, $j = 1, \dots,
N = L/2$. More generally, for every $N = 1, \dots, L/2$,
the lowest energy state in ${\cal H}_{L,L/2-N}$, i.e.,
for fixed magnetization $\mathfrak{m} = 1/2 - N/L$, has
$n_j = 1, \dots, N$. This was shown by Yang and Yang
\cite{YaYa66b} in the context of the more general XXZ
model (see below). For the Heisenberg model it had
been stated by Bethe \cite{Bethe31} that the lowest energy
state in ${\cal H}_{L,L/2-N}$ involves only real Bethe
roots and it had been conjectured by Hulth\'en
\cite{Hulthen38} that the ground state corresponds to
the above choice of integers.

Physical observables in the lowest-energy states at
fixed $\mathfrak{m}$ are calculated as sums of the
form $\frac{1}{L} \sum_{j=1}^N f(\la_j)$. Examples are
the total energy (\ref{eev}) and momentum (\ref{pev}),
but also ground-state correlation functions of the finite-length
chain \cite{DGHK07} are multiple sums of this form. As
in the case of free particles we expect simplifications
in the thermodynamic limit $L \rightarrow \infty$,
taken in such a way that $N$ depends on $L$ and $\lim_{L
\rightarrow \infty} N/L = D \in [0,1/2]$. The restriction
on $D$ is equivalent to $\mathfrak{m} = \bigl(\2 - D\bigr)
\in [0,1/2]$ which must be the case for Bethe Ansatz
eigenstates.  In the thermodynamic limit the `quantum numbers'
$(2j - N - 1)/2L$ on the right hand side of (\ref{logbae})
become equi-distributed variables $n$, densely filling the
interval $[-D/2, D/2]$. Assuming that, as a consequence
of (\ref{logbae}), the Bethe roots $\la_j$ get as well
continuously distributed with a distribution function
$\r (\la|q)$ and symmetric support $(-q, q)$, we expect
that observables characterized by (sufficiently
well-behaved) functions $f(\la)$ have the thermodynamic
limit
\begin{equation} \label{condens}
     \lim_{L \rightarrow \infty} \frac{1}{L}
        \sum_{j=1}^N f(\la_j) = \int_{-q}^q \rd \la \: \r (\la|q) f(\la) \epp
\end{equation}
In particular,
\begin{equation} \label{dq}
     \lim_{L \rightarrow \infty} \sum_{j=1}^N \frac{1}{L} = D
        = \int_{-q}^q \rd \la \: \r (\la|q) \epc
\end{equation}
while (\ref{logbae}) turns into
\begin{equation} \label{prelin}
     \frac{1}{\p} \arctg(2\la) = n(\la)
        + \int_{-q}^q \frac{\rd \m}{\p} \: \r(\m|q) \arctg(\la - \m) \epp
\end{equation}
Comparing this back with (\ref{logbae}) we see that
$\rd \la \, \r (\la|q) = \rd n(\la)$. Hence, differentiating
(\ref{prelin}), we obtain an integral equation for the
root density function,
\begin{equation} \label{rholin}
     \r (\la|q) = \frac{2}{\p (1 + 4 \la^2)}
        - \int_{-q}^q \frac{\rd \m}{\p} \: \frac{\r(\m|q)}{1 + (\la - \m)^2} \epp
\end{equation}

The system of equations (\ref{dq}), (\ref{rholin}) was first
considered by Hulth\'en. Hulth\'en managed to solve the
integral equation (\ref{rholin}) explicitly for $q = + \infty$
and observed that in this case $D = 1/2$, $\mathfrak{m} = 0$.
Assuming uniqueness of his solution he concluded that it
pertains to the ground state. Then he obtained the ground
state energy per lattice site $e = \lim_{L \rightarrow \infty} E/L$
from (\ref{eev}) and (\ref{condens}). This is actually not
difficult, if we use Fourier transformation and the convolution
theorem. The result is $e = - J \ln 2$. Later Yang and
Yang \cite{YaYa66c} proved that the system (\ref{dq}),
(\ref{rholin}) has indeed a unique solution $\rho(\la|q)$
for any $D \in [0,1/2]$. However, the fact that the limit on
the left hand side of (\ref{condens}) exists and equals the
right hand side of this equation was only proved much later
by K. K. Kozlowski in \cite{Kozlowski18b}.

\subsubsection*{Beyond the Heisenberg model}
After the works of Bethe and Hulth\'en followed a long
period of silence. A first generalization of Bethe's work was
obtained by R. Orbach in 1958 \cite{Orbach58}. Orbach observed
that the Heisenberg-Ising or XXZ chain,
\begin{equation} \label{hxxz}
     H_{XXZ} = \sum_{j=1}^L \bigl(s_j^x s_{j+1}^x + s_j^y s_{j+1}^y
                  + \D (s_j^z s_{j+1}^z - \tst{\4})\bigr) \epc
\end{equation}
$\D \in {\mathbb R}$, can be treated in a very similar way as
the Heisenberg chain, $\D = 1$. Since $[H_{XXZ},S^z] = 0$, we
can start with the $S^z$ eigenbasis ${\cal B}_z$ and with a
Bethe wavefunction (\ref{ba}) as above and obtain similar
results. Only the introduction of the appropriate rapidity
variables requires more thought. They were introduced in
subsequent work in \cite{Walker59}. In these variables the
Bethe Ansatz equations for the XXZ chain take the form
\begin{equation} \label{baexxz}
	\biggl(\frac{\sh(\la_j - \frac{\i \g}{2})}{\sh(\la_j + \frac{\i \g}{2})}\biggr)^L
	= - \prod_{k=1}^N \frac{\sh(\la_j - \la_k - \i \g)}{\sh(\la_j - \la_k + \i \g)} \epc
\end{equation}
where $j = 1, \dots, N$, and $\g$ is such that $\D = \cos(\g)$.

E. H. Lieb and W. Liniger in 1963 \cite{LiLi63} applied the
Bethe Ansatz to a very different kind of model, $N$ Bosons
on a one-dimensional ring of length $L$ interacting pairwise
via a repulsive delta-function potential. The Hamiltonian
takes the form
\begin{equation}
     H_{Bg} = - \sum_{j=1}^N \6_{x_j}^2 + 2c \sum_{1 \le j < k \le N} \de(x_j - x_k) \epp
\end{equation}
The stationary Schr\"odinger equation with this Hamiltonian
can be written as a wave equation for free particles plus
conditions for the scattering of pairs of particles plus
periodic boundary conditions. These conditions can be seen
as a continuous version of Eqs.~(\ref{waveeqn})-(\ref{period}).
Hence, it is not surprising that they are solved by Bethe's
Ansatz (\ref{ba}), provided that the Bethe Ansatz equations
\begin{equation} \label{baebose}
     \re^{\i k_j L} = - \prod_{\ell=1}^N \frac{k_j - k_\ell + \i c}{k_j - k_\ell - \i c} \qd
\end{equation}
hold for $j = 1, \dots, N$. It is remarkable that in this case
the quasi-momenta $k_j$ can be identified with the rapidity
variables.

Again much could be said about the interpretation of the Bethe
Ansatz solutions of the XXZ chain and of the Bose gas model.
We restrict ourselves, however, to remarking that the
XXZ chain and the Bose gas model have most frequently appeared
in numerous physical applications and are most probably the
best studied models the Bethe Ansatz has been applied to.
A peculiarity of the Bose gas model is that its solution sets
$\{k_j\}$ can be shown to be real \cite{YaYa69} and
in one-to-one correspondence with sets of integers $\{n_j\}
\in {\mathbb Z}$. This makes the analysis of various physical
quantities for the model much easier as compared to the spin
chains. The Bose gas model was the first model, whose excitations
were successfully studied on the basis of the Bethe Ansatz
solution \cite{Lieb63} and also the first model of many interacting
quantum particles for which the free energy was exactly calculated
\cite{YaYa69}.

\subsubsection*{Excited state of antiferromagnetic spin chains}
Excitations of the Heisenberg and Heisenberg-Ising spin
chains over their antiferromagnetic ground states were
studied on the basis of the Bethe Ansatz starting with
the work of J.~des Cloizeaux and J. J. Pearson \cite{ClPe62}.
These authors considered a family of excitations of the
Heisenberg chain involving only real Bethe roots. They
interpreted their result, in analogy with the magnons in
the ferromagnetic case, as spin-one excitations. Much later
it was understood \cite{FaTa81} that a more appropriate 
description of the excitations over the antiferromagnetic
ground state is in terms of spin-$\2$ excitations called
spinons. Spinons can only be created in pairs. These pairs
form scattering states and split in a singlet and a triplet,
which are degenerate. The singlet involves non-real Bethe
roots. The analysis in \cite{FaTa81} still relied on
problematic ingredients like strings. A more satisfactory
analysis of the elementary excitations over the antiferromagnetic
ground state, also for the XXZ chain, was developed in the works
\cite{Woynarovich82c,ViWo84,BVV83,DeLo82}, where so-called
higher-level Bethe equations were introduced. These can
nowadays be derived more convincingly by means of non-linear
integral equations \cite{KBP91} for the so-called counting
function (see e.g.~\cite{DGKS15a}).

\subsection{Conclusions on early Bethe Ansatz}
We have tried to give a detailed and technical description of
Bethe's work on the Heisenberg chain and its early extensions.
Our intention was to explain what the term Bethe Ansatz means
in a narrow sense and which questions emanated from the
original work. The Bethe Ansatz provides a large set
of eigenfunctions and eigenvalues of the Heisenberg
Hamiltonian and of other Hamiltonians that have been
studied later, parameterized in terms of solutions of
the Bethe Ansatz equations (Eqs.~(\ref{bae}) in case of
the Heisenberg chain). They can be used to study the
ground state properties and the excitations over the
ground state in the thermodynamic limit. However, the
mathematical questions connected with the analysis
of the Bethe Ansatz solutions are sometimes hard, and not
all of them have been answered until the present day.
These questions triggered many fruitful ideas which have
become part of what is nowadays understood under the term
Bethe Ansatz in a broader sense (for part of these ideas
see below). Before moving on with the discussion of a
few of the later developments let us summarize the difficulties
encountered so far.
\begin{enumerate}
\item
The completeness of the Bethe Ansatz is hard to establish
from the Bethe Ansatz equations alone. It needs advanced
methods. It has been established for the Heisenberg chain
\cite{MTV09} and for some other models, but remains a
challenge in the general case.
\item
The classification of solutions involving non-real Bethe
roots is tricky. A naive use of ideal strings gives incorrect
results, e.g., for the excitations over the antiferromagnetic
ground state. On the other hand the so-called string
hypothesis lead to a description of the thermodynamics of
the Heisenberg chain \cite{Takahashi71a,Gaudin71} that could be
confirmed by independent means. In any case, the string
hypothesis is an ingredient that should be avoided in more
serious mathematical studies.
\item
In general there are mathematical issues with the thermodynamic
limit, such as the condensation property which has been
established for the Heisenberg chain in \cite{Kozlowski18b}.
\item
Another point that remained unclear until the end of the
60s is the question why some models are tractable by
Bethe Ansatz and others not, or what might be the mathematical
structure behind the Bethe Ansatz. The latter question found
its answer in the connection of the Bethe Ansatz solvable
quantum chains with vertex models that will be explained in
the next section.
\end{enumerate}

\section{Vertex models}
\subsection{The six- and eight-vertex models}
\subsubsection*{The six-vertex model and its partition function}
\begin{figure}
\begin{center}
\includegraphics[width=.96\textwidth]{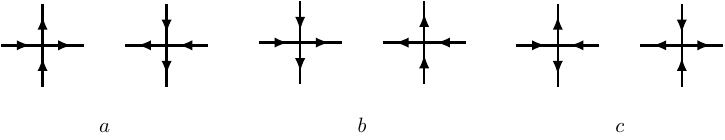}
\caption{\label{fig:six-vtxs} The six vertex configurations
compatible with the ice rule.}
\end{center}
\end{figure}

The story of the Bethe Ansatz took an unexpected turn
in 1967, when E. H. Lieb managed to apply it \cite{Lieb67,Lieb67c}
to the solution of a longstanding problem in statistical
mechanics. He exactly calculated the entropy of a two-dimensional
version of a model for the ground-state degeneracy of ice.
In real ice the oxygen atoms form a regular lattice with
four hydrogen bonds to their oxygen neighbours, and
the hydrogen atoms are placed on the bonds in such a way that
two of them are closer to and two of them are farther away from
the central oxygen atom. With this so-called ice rule there
are $\bigl(\begin{smallmatrix}4\\2\end{smallmatrix}\bigr) = 6$
local bond configurations. The configurations can be depicted
by placing arrows pointing toward a lattice point for the
closer atoms and arrows pointing away from the lattice points
for the farther ones. The ice rule then says that there are two
arrows pointing in and two arrows pointing out around every
lattice point. A lattice point together with the four bonds
connecting it to its neighbours is called a vertex. In the ice
model there are six different local vertex configurations (see
Fig.~\ref{fig:six-vtxs}). Lieb solved the problem of counting
the number of configurations obeying the ice rule per lattice
site in a rectangular lattice with periodic boundary conditions,
when its size goes to infinity.

More generally we may assign an energy or a local Boltzmann
weight to every of the six vertex configurations (see
Fig.~\ref{fig:six-vtxs}). If we do it in such a way that
configurations that are related by the reflection of all
arrows have the same energy, we remain with three different
Boltzmann weights $a$, $b$, $c$ and we have defined the (symmetric)
six-vertex model \cite{Sutherland67}. The ice model is then
contained as the special case $a = b = c = 1$. If we denote
a configuration of arrows on an $L \times M$ periodic square
lattice that is compatible with the ice rule by $\s$, and the
numbers of vertices of type $a$, $b$, or $c$ in a given
configuration $\s$ by $n_a (\s)$, $n_b (\s)$, or $n_c (\s)$,
then the partition function of the six-vertex model under
periodic boundary conditions becomes
\begin{equation}
     Z_{L, M} (a, b, c) = \sum_\s a^{n_a (\s)} b^{n_b (\s)} c^{n_c (\s)} \epp
\end{equation}

\subsubsection*{The six-vertex model transfer matrix}
A decisive step in Lieb's work \cite{Lieb67} was to represent
the partition function by means of an appropriate transfer
matrix \cite{KrWa41a}. The construction can be described as
follows. Define $R \in \End \bigl({\mathbb C}^2 \otimes
{\mathbb C}^2\bigr)$ by setting
\begin{equation} \label{relms}
     R_{++}^{++} = R_{--}^{--} = a \epc \qd
     R_{+-}^{+-} = R_{+-}^{+-} = b \epc \qd
     R_{-+}^{+-} = R_{+-}^{-+} = c
\end{equation}
and $R^{\a \be}_{\g \de} = 0$ in the remaining ten cases. Let
\begin{equation}
     T_0 = R_{0, L} \dots R_{0, 1} \in \End \bigl({\mathbb C}^2\bigr)^{\otimes (L+1)} \epp
\end{equation} 
Then $\tr_0 T_0$ is called the transfer matrix of the six-vertex
model, and
\begin{equation}
     Z_{L, M} (a, b, c) = \tr_{1, \dots, L} \Bigl\{\bigl(\tr_0 T_0\bigr)^M\Bigr\} \epp
\end{equation}
If $\La_0 (a,b,c)$ is the largest eigenvalue of the transfer matrix
$\tr_0 T_0$, then
\begin{equation}
     \lim_{M \rightarrow \infty} \frac{1}{M} \ln Z_{L, M} (a, b, c)
        = \ln \La_0 (a,b,c) \epp
\end{equation}
Hence, a single eigenvalue of the transfer matrix determines the
asymptotics of the partition function for $M \rightarrow \infty$.
Lieb's famous result on the entropy of `square ice' is
\begin{equation}
     \lim_{L \rightarrow \infty} \lim_{M \rightarrow \infty}
     \frac{1}{LM} \ln Z_{L, M} (1, 1, 1) =
        \lim_{L \rightarrow \infty} \frac{1}{L} \ln \La_0 (1,1,1) =
	\frac{3}{2} \ln \biggl(\frac{4}{3}\biggr) \epp
\end{equation}

In order to obtain this result Lieb diagonalized the transfer
matrix using the Bethe Ansatz (\ref{ba}). He found that in this
case the wave functions are exactly the same as those for the
Heisenberg-Ising chain (\ref{hxxz}) that had been obtained by
Orbach \cite{Orbach58} and were studied by Yang and Yang
\cite{YaYa66b}. Then McCoy and Wu in \cite{McWu68} in a more
general case of a six-vertex model showed that the transfer
matrix commutes with the Hamiltonian (\ref{hxxz}).

\subsubsection*{Baxter's early work}
Finally, a breakthrough in the understanding of the Bethe
Ansatz solvable models and a groundbreaking generalization
of the method was achieved by R. J. Baxter in 1971
\cite{Baxter71a,Baxter71b,Baxter71c,Baxter72}. His works
radically changed the view of the Bethe Ansatz solvable models
and laid the foundations for many subsequent developments.
\begin{enumerate}
\item
Baxter considered an inhomogeneous six-vertex model and found
that the Bethe Ansatz method works if two transfer matrices
corresponding to different rows with different parameters
commute.
\item
He found that a `local commutativity condition' on the
`$R$-matrices' defined in (\ref{relms}) guarantees the
commutativity of the transfer matrices. He proved it for
the six-vertex model and for the more general eight-vertex
model which has non-vanishing Boltzmann weights $R^{++}_{--}
= R^{--}_{++} = d$ in addition to (\ref{relms}). His
condition is nowadays known as the Yang-Baxter equation.
It had appeared before in the different context of
fatorizable multi-particle scattering \cite{Yang68}.
\item
Baxter found a proper parameterization of the $R$-matrices
related to the uniformization of a complex curve. In
this parameterization $R$ is carrying a `spectral parameter'
$\la$ and further `deformation parameters' which play a
different role. Transfer matrices composed of $R$-matrices
satisfying the Yang-Baxter equation commute if they have
the same deformation parameters but arbitrary values
of the spectral parameters. Hence, the commuting transfer
matrices have joint eigenvectors that do not depend on
the spectral parameter. The latter appears only in the
eigenvalues, whence its name.
\item
Baxter found that the logarithmic derivative of the
six-vertex model transfer matrix gives the Heisenberg-Ising
Hamiltonian, whereas the logarithmic derivative of the
eight-vertex model transfer matrix that of the totally
anisotropic XYZ spin chain \cite{Baxter71c}.
\item
Baxter found that the Bethe Ansatz equations can be
replaced by a functional equation, now called the Baxter
equation or Baxter's $TQ$ equation.
\item
He visionary postulated the existence of a `$Q$ operator'
from the $TQ$ equation which remained the subject of
contestation for a while, but later became well rooted
in the representation theory of quantum groups \cite{BLZ97,%
Korff05,BLMS10}. Meanwhile it has become an important part
of the modern theory of Yang-Baxter integrable models
\cite{NirRaz14}.
\end{enumerate}
\subsubsection*{The Yang-Baxter equation}
We shall briefly illustrate some of the above points.
In order to keep our discussion simple we restrict
ourselves to the six-vertex model. Consider the
reparameterization $a = \r \sh(\la + \h)$, $b = \r \sh(\la)$,
$c = \r \sh(\h)$ of the Boltzmann weights in (\ref{relms}).
This can be solved for $\la$, $\h$, $\r$ if $a, b, c \ne
0$ and if $a \pm b \ne -c , c$. We shall write $R = R(\la)$
and keep the dependence on $\h$ and $\r$ implicit. Then
it is not difficult to see that
\begin{equation} \label{ybe}
     R_{1,2} (\la - \m) R_{1,3} (\la - \n) R_{2,3} (\m - \n) =
	R_{2,3} (\m - \n) R_{1,3} (\la - \n) R_{1,2} (\la - \m)
\end{equation}
which is the Yang-Baxter equation mentioned above. The
$R$-matrix (\ref{relms}) has the important property that
\begin{equation} \label{reg}
       R(0) = \r \sh(\h) P \epc
\end{equation}
where $P$ is the transposition matrix (\ref{transmat}).

With every set $\{\x_1, \dots, \x_L\} \subset {\mathbb C}$
of `inhomogeneity parameters' we associate an
inhomogeneous `monodromy matrix'
\begin{equation} \label{defmono}
     T_0 (\la) = R_{0, L} (\la - \x_L) \dots R_{0, 1} (\la - \x_1)
        \in \End \bigl({\mathbb C}^2\bigr)^{\otimes (L+1)} \epp
\end{equation}
We interpret it as an object acting on $L$ sites representing
the physical degrees of freedom of a lattice times one
auxiliary site with index $0$. Extending the system by
two auxiliary sites $0$, $0'$ we can define two monodromy
matrices $T_0 (\la)$ and $T_{0'} (\m)$. As a consequence of
the Yang-Baxter equation they satisfy the relation
\begin{equation} \label{yba}
     R_{0, 0'} (\la - \m) T_0 (\la) T_{0'} (\m) =
        T_{0'} (\m) T_0 (\la) R_{0, 0'} (\la - \m) \epp
\end{equation}
If we multiply by the inverse of $R_{0, 0'} (\la - \m)$ from
the left or from the right and take the trace in the tensor
product of spaces $0$ and $0'$, we see that
\begin{equation}
     [\tr_0 T_0 (\la), \tr_0 T_0 (\m)] = 0 \epp
\end{equation}
Two transfer matrices with different values of the spectral
parameter commute as a consequence of the Yang-Baxter equation
(\ref{ybe}).

\subsubsection*{Relation between six-vertex model and XXZ quantum spin chain}
In the homogeneous case, $\x_j = 0$, $j = 1, \dots, L$,
Eq.~(\ref{reg}) implies that
\begin{equation}
     \tr_0 T_0 (0) = \r^L \sh^L (\h) \hat U \epc
\end{equation}
where $\hat U$ is the shift operator (\ref{defu}). Using
this result we also see that
\begin{equation} \label{hamtm}
     \bigl(\tr_0 T_0 (0)\bigr)^{-1} \tr_0 T_0' (0) =
        \frac{2}{J \sh (\h)} \bigl(H_{XXZ} + \D L/2\bigr) \epc
\end{equation}
where we have identified $\D = \ch(\h)$. In other words, up to
a shift by a constant and up to a change of the normalization
the Hamiltonian of the XXZ chain is equal to the logarithmic
derivative of the homogeneous six-vertex model transfer matrix.
This explains the observation of McCoy and Wu, that $H_{XXZ}$
commutes with the transfer matrix of the six-vertex model.
We also conclude that $H_{XXZ}$ commutes with all higher logarithmic
derivatives of the six-vertex model transfer matrix. Thus,
it is only one of many mutually commuting independent operators.
This gives an answer to the question what might be special
about the Heisenberg-Ising Hamiltonian.

As we shall see, the inhomogeneous transfer matrix can be equally
well diagonalized by the Bethe Ansatz. This becomes particularly
transparent within the algebraic Bethe Ansatz approach to be
discussed below.

\subsection{Nested Bethe Ansatz}
An important development taking place in the late 60s was the
invention of the nested Bethe Ansatz method in the works of M. Gaudin
\cite{Gaudin67} and C. N. Yang \cite{Yang67} and its application
to the Hubbard model by E. H. Lieb and F. Y. Wu \cite{LiWu68}
and to the higher-rank isotropic spin chains by B.~Sutherland
\cite{Sutherland75}. In these works the Bethe Ansatz was
generalized to deal with models with more complicated local
Hilbert spaces.

\subsubsection*{The one-dimensional Hubbard model}
Most important in applications among the above cited works
is probably the Hubbard model, a one-band electronic model
with nearest-neighbour hopping defined by the Hamiltonian
\cite{LiWu68}
\begin{equation} \label{htu}
     H_u = - \sum_{j=1}^L \sum_{a = \auf, \ab}
	   (c_{j, a}^\+ c_{j+1, a}^{} + c_{j+1, a}^\+ c_{j, a}^{})
	   + u \sum_{j=1}^L (1 - 2 n_{j \auf})(1 - 2 n_{j \ab}) \epp
\end{equation}
Here $c_{j, a}^\+$ and $c_{j, a}^{}$ are creation and annihilation
operators of electrons of spin $a$ ($a = \auf$ or $a = \ab$) localized
in an orbital at site $j$ of a one-dimensional lattice, and $n_{j,a} =
c_{j, a}^\+ c_{j, a}^{}$. The operators $c_{j, a}^\+$ and 
$c_{j, a}^{}$ satisfy the canonical anticommutation relations
\begin{subequations} \label{comab}
\begin{align} \label{coma}
     & \{c^{}_{j, a}, c^{}_{k, b}\} = \{c_{j, a}^\+, c_{k, b}^\+\} = 0
        \epc \\
     & \{c^{}_{j, a}, c_{k, b}^\+\} = \de_{jk} \de_{ab} \label{comb}
\end{align}
\end{subequations}
for $j, k = 1, \dots, L$ and $a, b = \auf, \ab$. The parameter
$u$ is real and determines the interaction strength. Periodic
boundary conditions on the operators, $c_{L+1, a} = c_{1, a}$
are understood. The Hubbard model \cite{Hubbard63}, defined on
one-, two- and three-dimensional crystal lattices, is a key
model in the theory of strongly correlated electron systems
in condensed matter physics.

\subsubsection*{Space of states and Bethe Ansatz wave function}
The creation operators $c_{j, a}^\+$ generate the space of
states ${\cal H}^{(L)}$ of the Hubbard model by their action
on a pseudo vacuum $|0\>$ defined by the condition
\begin{equation} \label{vacuum}
     c_{j, a} |0\> = 0 \epc \qd j = 1, \dots, L \epc
			    \qd a = \auf, \ab \epp
\end{equation}
We introduce row vectors of electron and spin coordinates, $\xv =
(x_1, \dots, x_N)$ and $\av = (a_1, \dots, a_N)$ with $x_j \in
\{1, \dots, L\}$ and $a_j = \auf, \ab$. The space of states of the
Hubbard model is spanned by all linear combinations of the so-called
Wannier states
\begin{equation} \label{basisstate}
     |\xv, \av \> = c_{x_N, a_N}^\+ \dots c_{x_1, a_1}^\+ |0\> \epp
\end{equation}
If the coordinates $x_j$ and $a_k$ are appropriately ordered,
these states form a basis in which the Hamiltonian $H_u$ is block
diagonal, since it preserves the particle number and the
$z$~component of the total spin.

The solution of the eigenvalue problem within the blocks proceeds
along similar lines as for the Heisenberg chain. One first
translates the eigenvalue problem of the Hamiltonian into a
set of relations for a Bethe Ansatz wave function, which
look similar to those in Lemma~\ref{lem:firstqh}. The
difference is that the amplitudes in the wave function
are now coordinates of a vector. The wave equation, together
with the requirement that the wave function is antisymmetric
and satisfies periodic boundary conditions, translates into an
eigenvalue equation for this vector that is equivalent to the
eigenvalue problem of the transfer matrix of an inhomogeneous
six-vertex model. The latter can be solved by recourse to
the work of Lieb \cite{Lieb67b} and Sutherland \cite{Sutherland67},
or nowadays more conveniently by the algebraic Bethe Ansatz (see
below). For more details the reader is referred to the monograph
\cite{thebook}.

\enlargethispage{-3ex}

\subsubsection*{Bethe Ansatz solution of the Hubbard model}
\begin{theorem} \cite{LiWu68,thebook}.
\begin{enumerate}
\item
In a block with $N$ electrons and $M$ down spins the eigenstates
of the one-dimensional Hubbard model are characterized by two
row vectors $\kv = (k_1, \dots, k_N)$ and $\lav = (\la_1, \dots,
\la_M)$ of quantum numbers, for which $2M \le N \le L$.
\item
The Bethe Ansatz eigenvectors can be represented as
\begin{equation} \label{state}
     |\ps_{\kv, \lav}\> = \frac{1}{N!} \sum_{x_1, \dots, x_N = 1}^L\
	      \sum_{a_1, \dots, a_N = \uparrow, \downarrow}
	      \ps (\xv; \av|\kv; \lav) |\xv, \av\> \epc
\end{equation}
where $\ps (\xv; \av|\kv; \lav)$ is the $N$-particle Bethe ansatz wave
function. The latter depends on the relative ordering of the coordinates
$x_j$. Any ordering is assigned to a permutation $Q \in {\mathfrak S}^N$
through the inequality
\begin{equation} \label{sectorq}
     1 \le x_{Q(1)} \le x_{Q(2)} \le \dots \le x_{Q(N)} \le L \epp
\end{equation}
The inequality (\ref{sectorq}) divides the configuration space of $N$
electrons into $N!$ sectors, which can be labeled by the permutations
$Q$. In sector $Q$ the Bethe ansatz wave functions take the form
\begin{equation} \label{wwf}
     \ps(\xv; \av| \kv; \lav) = \sum_{P \in {\mathfrak S}^N} \sign(PQ)
                                \<\av Q|\kv P, \lav\> \,
				\re^{\i \<\kv P, \xv Q\>}
\end{equation}
with spin dependent amplitudes $\<\av Q|\kv P, \lav \>$.
\item
The amplitudes take the form of the Bethe Ansatz wave functions
of an inhomogeneous Heisenberg spin chain, i.e.,
\begin{equation} \label{wswf}
     \<\av Q|\kv P, \lav\> = \sum_{R \in {\mathfrak S}^M} A(\lav R)
			     \prod_{\ell=1}^M
			     F_{\kv P} (\la_{R(\ell)}; y_\ell) \epc
\end{equation}
where
\begin{equation} \label{wwfsf}
     F_{\kv} (\la; y) = \frac{2 \i u}{\la - \sin k_y + \i u}
		    \prod_{j=1}^{y-1} \frac{\la - \sin k_j - \i u}
		                           {\la - \sin k_j + \i u} \epc
\end{equation}
and
\begin{equation} \label{wwfsa}
     A(\lav) = \prod_{1 \le m < n \le M}
	\frac{\la_m - \la_n - 2 \i u}{\la_m - \la_n} \epp
\end{equation}
In the above equations $y_j$ denotes the position of the $j$th down
spin in the sequence $a_{Q(1)}, \dots, a_{Q(N)}$. The $y$'s are thus
`coordinates of down spins on electrons'. If the number of down spins
in the sequence $a_{Q(1)}, \dots, a_{Q(N)}$ is different from $M$, the
amplitude $\<\av Q|\kv P, \lav\>$ vanishes.
\item
The quantum numbers $k_j$, $j = 1, \dots, N$, and $\la_\ell$, $\ell =
1, \dots, M$, may be non-real. They are called charge momenta
and spin rapidities. They solve the Bethe Ansatz equations
\begin{align} \label{bak}
     & \re^{\i k_j L} = \prod_{\ell=1}^M
                        \frac{\la_\ell - \sin k_j - \i u}
			     {\la_\ell - \sin k_j + \i u}
			 \epc \qd j = 1, \dots, N \epc \\ \label{bas}
     & \prod_{j=1}^N \frac{\la_\ell - \sin k_j - \i u}
                          {\la_\ell - \sin k_j + \i u} =
       \prod_{m=1 \atop m \ne \ell}^M \frac{\la_\ell - \la_m - 2 \i u}
                                           {\la_\ell - \la_m + 2 \i u}
			 \epc \qd \ell = 1, \dots, M
\end{align}
which in this case are also called the Lieb-Wu equations.
Note that the restrictions $2M \le N \le L$ are imposed on
the numbers of charge momenta and spin rapidities.
\item
The states (\ref{state}) are joint eigenstates of the Hubbard
Hamiltonian (\ref{htu}) and the corresponding momentum operator
with eigenvalues
\begin{equation} \label{enmom}
     E = - 2 \sum_{j=1}^N \cos k_j + u (L - 2N) \epc \qd
     P = \Biggl[ \sum_{j=1}^N k_j \Biggr] \mod 2\p \epp
\end{equation}
\item
The states (\ref{state}) are highest weight states with respect
to the total spin and with respect to another $\mathfrak{sl}_2$
symmetry called the $\h$-pairing symmetry \cite{EKS92b}.
\end{enumerate}
\end{theorem}

\subsection{Algebraic Bethe Ansatz}
\subsubsection*{Connection with classical integrable evolution equations}
Another variant was added to the analysis of Bethe Ansatz
solvable models by L. D. Faddeev and his school. They started
out from the analysis of integrable classical evolution equations
\cite{GGKM67,AKNS73,ZaSh71} which had been interpreted as
integrable Hamiltonian systems \cite{ZaFa71} and managed to
lift the inverse scattering transform \cite{GGKM67} that
had been invented for solving the classical models to the
quantum level \cite{Sklyanin80,Sklyanin78b,STF79}. Sklyanin
observed \cite{Sklyanin80} that the structure of the fundamental
Poisson brackets of the classical `transition coefficients' is 
encoded in a classical $R$-matrix satisfying a `classical
Yang-Baxter equation'.

In \cite{STF79} the authors treated the quantum Sine-Gordon
model. They showed that the commutation relations of the
quantum analogues of the `transition coefficients' obey the
relations (\ref{yba}) with the same $R$-matrix of the
six-vertex model. This observation placed the $R$-matrix,
and with it the Yang-Baxter equation, in the center of the
theory of integrable systems. While Baxter had used the
relations (\ref{yba}), which are nowadays often called
Yang-Baxter algebra relations, only for showing that two
transfer matrices with different spectral parameter commute
with each other, Sklyanin, Takhtadjan and Faddeev used
the same relations for a simple algebraic construction of
the Bethe vectors which they later called the algebraic Bethe
Ansatz.

Clearly, the steps that lead from the Yang-Baxter equation
(\ref{ybe}) and the so-called regularity condition (\ref{reg})
to the derivation of the Yang-Baxter algebra relations
(\ref{yba}) and to the (local) Hamiltonian (\ref{hamtm})
are very general. Any solution of the Yang-Baxter equations
that satisfies (\ref{reg}) induces similar structures and
in this sense defines a solvable or `Yang-Baxter integrable'
lattice model. With a little more effort this idea can
be generalized to more general classes of models. This
insight \cite{KuSk80} initiated a quest for a solution
theory of the Yang-Baxter equation (\ref{ybe}) which 
lead to the advent of quantum groups \cite{Drinfeld85,%
Jimbo85} and established a connection between the
Bethe Ansatz solvable models and the representation theory
of quantum groups that carries on to be fruitful.

\subsubsection*{The algebraic Bethe Ansatz for the six-vertex model}
Let us inspect the algebraic Bethe Ansatz for the six-vertex
model. For simplicity we consider the homogeneous case and
set $\x_j = \frac{\h}{2}$ for $j = 1, \dots, L$ in (\ref{defmono}).
We first of all represent the monodromy matrix (\ref{defmono})
as a $2 \times 2$ matrix in the auxiliary space
`$0$',
\begin{equation}
     T_0 (\la) = \begin{pmatrix} A(\la) & B(\la) \\ C(\la) & D(\la) \end{pmatrix}_0 \epp
\end{equation}
Here $A(\la), \dots, D(\la) \in \End \bigl({\cal H}_L\bigr)$.
Expressed in terms of these operators the transfer matrix
takes the form $\tr_0 T_0 (\la) = A(\la) + D(\la)$.
The local relation $[R(\la),s^z \otimes I_2 + I_2 \otimes s^z] = 0$
implies that
\begin{subequations}
\begin{align}
     & [A(\la), S^z] = [D(\la),S^z] = 0 \epc \qd \\[1ex]
     & S^z B(\la) = B(\la) (S^z - \id) \epc \qd
       S^z C(\la) = C(\la) (S^z + \id) \epc
\end{align}
\end{subequations}
meaning that $A(\la)$ and $D(\la)$ preserve the $z$ component
of the spin, while $B(\la)$ lowers it by $1$ and $C(\la)$
raised it by $1$. It is further not difficult to calculate
the action of $B(\la)$ on the pseudo-vacuum state (\ref{pseudo}),
\begin{equation}
     B(\la) |0\> =
        \frac{\r^L}{\sh(\la - \frac{\h}{2}) \sh(\la + \frac{\h}{2})}
	\sum_{x=1}^L \sh(\la - \tst{\frac{\h}{2}})^x
	   \sh(\la + \tst{\frac{\h}{2}})^{L - x + 1} |(x)\> \epp
\end{equation}
Comparison with (\ref{wavefunren}) shows that the coefficients
under the sum are equal to the off-shell Bethe wave function
for a single overturned spin (in the XXZ case). This finding
holds more generally \cite{IKR87}. Up to a change in normalization
a product of `B-operators' applied to the pseudo vacuum generates
the XXZ version of the off-shell Bethe states (\ref{wavefunren}),
(\ref{bvoffshell}). Performing an algebraic Bethe
Ansatz thus means to show that a product of $B$-operators
applied to the pseudo vacuum generates an eigenstate
of $A(\la) + D(\la)$, if appropriate Bethe Ansatz equations
are satisfied. This can be achieved by means of the
Yang-Baxter algebra~(\ref{yba}).

The Yang-Baxter algebra relations (\ref{yba}) are a set of
16 quadratic relations for the mono\-dromy matrix elements
$A(\la), \dots, D(\la)$ including, in particular, the relations
\begin{subequations}
\label{relaba}
\begin{align}
     B(\la) B(\m) & = B(\m) B(\la) \epc \\[1ex]
     A(\la) B(\m) & = \frac{\sh(\m - \la + \h)}{\sh(\m - \la)} B(\m) A(\la)
                      - \frac{\sh(\h)}{\sh(\m - \la)} B(\la) A(\m) \epc \\[1ex]
     D(\la) B(\m) & = \frac{\sh(\la - \m + \h)}{\sh(\la - \m)} B(\m) D(\la)
                      - \frac{\sh(\h)}{\sh(\la - \m)} B(\la) D(\m) \epp
\end{align}
\end{subequations}
For the algebraic Bethe Ansatz we further need the pseudo-vacuum
actions
\begin{equation} \label{vacact}
     A(\la)|0\> = a(\la) |0\> \epc \qd
     D(\la)|0\> = d(\la) |0\> \epc \qd
     C(\la)|0\> = 0 \epc
\end{equation}
where $a(\la) = \r^L \sh^L (\la + \frac{\h}{2})$,
$d(\la) = \r^L \sh^L (\la - \frac{\h}{2})$.

For subsets of ${\mathbb C}$ we shall use the
short-hand notations $\{\la\} = \{\la_j\}_{j=1}^{N+1}$,
$\{\la\}_\ell = \{\la_j\}_{j=1, j \ne \ell}^{N+1}$. We
further introduce the functions
\begin{equation}
     Q(\la|\{\n\}) = \prod_{\n \in \{\n\}} \sh(\la - \n) \epp
\end{equation}
The relations (\ref{relaba}) can be used iteratively together
with the vacuum action (\ref{vacact}) to calculate the action
of $A(\la)$ and $D(\la)$ on off-shell states
\begin{equation}
     {\mathbb B} (\{\la\}_\ell) =
        \prod_{j=1, j \ne \ell}^{N+1} B(\la_j) |0\> \epp
\end{equation}
Adding up the resulting expressions we obtain the action of
the transfer matrix on off-shell Bethe vectors in the form
\begin{multline} \label{leftoffshellaction}
     \tr_0 \{T_0 (\la_\ell)\}\, {\mathbb B} (\{\la\}_\ell) \\ =
        \sum_{j=1}^{N+1}
	   \frac{a(\la_j) Q(\la_j - \h|\{\la\}_\ell)
	         + d(\la_j) Q(\la_j + \h|\{\la\}_\ell)}{Q(\la_j|\{\la\}_j)}\,
		   {\mathbb B} (\{\la\}_j) \epp
\end{multline}
We can conclude with the following
\begin{theorem}
The off-shell vector ${\mathbb B} (\{\la\}_\ell)$ becomes an
eigenvector of the transfer matrix  $\tr_0 \{T_0 (\la_\ell)\}$ if
\begin{equation} \label{baeqform}
     a(\la_j) Q(\la_j - \h|\{\la\}_\ell) + d(\la_j) Q(\la_j + \h|\{\la\}_\ell) = 0
\end{equation}
for $\la_j \in \{\la\}_\ell$. The corresponding transfer
matrix eigenvalue is then
\begin{equation}
     \La (\la_\ell|\{\la\}_\ell) =
        \frac{a(\la_\ell) Q(\la_\ell - \h|\{\la\}_\ell)
	      + d(\la_\ell) Q(\la_\ell + \h|\{\la\}_\ell)}
	     {Q(\la_\ell|\{\la\}_\ell)} \epp
\end{equation}
\end{theorem}
Eqs.~(\ref{baeqform}) are nothing but the Bethe Ansatz
equations. They can be easily brought to the form 
(\ref{baexxz}). The logarithmic derivative of
the transfer matrix eigenvalue with respect to $\la_\ell$
at $\la_\ell = \frac{\h}{2}$ gives the corresponding energy
eigenvalue of the XXZ Hamiltonian (\ref{hxxz}). The
only modification that is required for the inhomogeneous
case (\ref{defmono}) is the replacement of the vacuum
expectation values $a(\la)$ and $d(\la)$ by
\begin{equation}
     a(\la) = \r^L \prod_{j=1}^L \sh(\la - \x_j + \h) \epc \qd
     d(\la) = \r^L \prod_{j=1}^L \sh(\la - \x_j) \epp
\end{equation}

\subsubsection*{Pairing between on- and off-shell Bethe vectors}
One of the main achievements of the algebraic Bethe
Ansatz, at least when applied to the six-vertex model,
is that it allows to derive a determinant formula for the
pairing (or `scalar product') of on- and off-shell
Bethe vectors \cite{Slavnov89,BeSl19} that turned out
to be utterly useful for the calculation of correlation
functions of local operators within the Bethe Ansatz
approach.

`Dual off-shell Bethe vectors' are defined as
\begin{equation}
     {\mathbb C} (\{\m\}_\ell) =
        \<0| \prod_{j=1, j \ne \ell}^{N+1} C(\m_j) \epc
\end{equation}
where $\<0|$ is the dual pseudo vacuum. They satisfy
a relation dual to (\ref{leftoffshellaction}),
\begin{multline} \label{rightoffshellaction}
     {\mathbb C} (\{\m\}_\ell) \tr_0 \{T_0 (\m_\ell)\} \\ =
	\sum_{j=1}^{N+1} {\mathbb C} (\{\m\}_j)
	   \frac{a(\m_j) Q(\m_j - \h|\{\m\}_\ell)
	         + d(\m_j) Q(\m_j + \h|\{\m\}_\ell)}{Q(\m_j|\{\m\}_j)} \epp
\end{multline}

Let us set $\{\m\}_{N+1} = \{\m\}$, $\m_{N+1} = \la_\ell$,
and let us assume that $\{\m\}$ satisfies the Bethe Ansatz
equations (\ref{baeqform}). Then
\begin{equation}
     {\mathbb C} (\{\m\}) \tr_0 \{T_0 (\la_\ell)\} =
        {\mathbb C} (\{\m\}) \La(\la_\ell|\{\m\}) \epp
\end{equation}
Setting $X^j = {\mathbb C} (\{\m\}) {\mathbb B} (\{\la\}_j)$ and
multiplying (\ref{leftoffshellaction}) by ${\mathbb C} (\{\m\})$
from the left we obtain the following set of linear
equations for the $X^j$,
\begin{equation} \label{linsys}
        \sum_{j=1}^{N+1}
	   \frac{a(\la_j) Q(\la_j - \h|\{\la\}_\ell)
	         + d(\la_j) Q(\la_j + \h|\{\la\}_\ell)}{Q(\la_j|\{\la\}_j)} X^j
		 = \La(\la_\ell|\{\m\}) X^\ell \epp
\end{equation}

This system can be solved for the $X^j$ \cite{BeSl19},
but, being homogeneous, only up to an overall normalization.
Fortunately, the latter drops out in applications. In
oder to state the result for ratios of pairings we introduce
the notations
\begin{subequations}
\begin{align}
     & e(\la) = \cth(\la) - \cth(\la + \h) \epc \qd
     K(\la) = \cth(\la - \h) - \cth(\la + \h) \epc \\[1ex]
     & \fa (\la|\{\m\}) = \frac{d(\la) Q(\la + \h|\{\m\})}{a(\la) Q(\la - \h|\{\m\})} \epp
\end{align}
\end{subequations}
With these notations we obtain the following
\begin{theorem} Normalized Slavnov formula \cite{Slavnov89}.
If $\{\m\}$ satisfies the Bethe Ansatz equations (\ref{baeqform}),
if ${\mathbb B} (\{\m\})$ is the corresponding on-shell Bethe
vector with dual ${\mathbb C} (\{\m\})$, and if ${\mathbb B} (\{\la\}_{N+1})$
is any off-shell Bethe vector, then
\begin{equation} \label{slavnov}
     \mspace{-6.mu}
     \frac{{\mathbb C} (\{\m\}) {\mathbb B} (\{\la\}_{N+1})}
          {{\mathbb C} (\{\m\}) {\mathbb B} (\{\m\})} = \\
     \biggl[\prod_{j=1}^N \frac{\La(\la_j|\{\m\})}{\La(\m_j|\{\m\})}\biggr]
     \frac{\det_N \Bigl\{\frac{e(\m_j - \la_k)}{1 + \fa(\la_k|\{\m\})} -
           \frac{e(\m_j - \la_k)}{1 + \fa^{-1} (\la_k|\{\m\})}\Bigr\}}
          {\det_N\Bigl\{\de^j_k - \frac{K(\m_j - \m_k)}{\fa'(\m_k|\{\m\})}\Bigr\}
	   \det_N\Bigl\{\frac{1}{\sh(\m_j - \la_k)}\Bigr\}}.
\end{equation}
\end{theorem}

At this point two more remarks might be appropriate. First
of all, although there is a general scheme how to connect
a solution of the Yang-Baxter equation with a vertex
model and with an associated Yang-Baxter algebra, there
is no general scheme how to efficiently construct off-shell
or on-shell Bethe vectors and, in general, no formulae like
(\ref{slavnov}) are known. Despite much progress in recent
years (e.g.\ in \cite{PRS15a,PRS15b}) a more efficient algebraic
Bethe Ansatz for the nested case and for the Hubbard model,
in particular, is yet to be developed. Second, the use of
$Q$-operators and functional equations makes it possible
to avoid the construction of Bethe vectors at all. The
method of $Q$-operators \cite{NirRaz14} is a method to solve
the transfer matrix eigenvalue problem without constructing
eigenvectors.

\section{Bethe Ansatz in quantum statistical mechanics}
One of the big promises of the Bethe Ansatz is that, one day,
it will provide us with rigorous exact solutions of the basic
problems of quantum statistical mechanics for interacting
quantum chains such as the Heisenberg-Ising model or the
Hubbard model. Those basic problems are the calculation of
the partition function and free energy per lattice site in
the thermodynamic limit and the calculation of static
and dynamic correlation functions in thermal equilibrium.

This programme is work in progress. As an extension of
\cite{YaYa69} a `thermodynamic Bethe Ansatz' has been
developed for many Yang-Baxter integrable models, much
of it in the works of M. Takahashi \cite{Takahashi99}. This
includes, in particular, work on the Heisenberg chain
\cite{Takahashi71a} and on the Hubbard model \cite{Takahashi72,%
Takahashi74}. The thermodynamic Bethe Ansatz relies on
the string hypothesis, and it typically involves the solution of
an infinite coupled system of nonlinear integral equations.
These features are problematic for a rigorous justification
as well as for the numerical calculation of thermodynamic
properties. An alternative approach, free of these shortcomings,
was developed based on the so-called quantum transfer matrix
formalism \cite{Suzuki85} which was adapted to the
realm of integrable models in a series of works
\cite{SAW90,Koma87, Koma89,DeVe92} in the late 80s and early
90s. It found its most efficient formulation in \cite{Kluemper93},
where it was applied to the XXZ chain. Later in \cite{JKS98c}
it was also successfully applied to the Hubbard model. The quantum
transfer matrix approach was rigorously justified for the XXZ chain
at sufficiently high temperatures in \cite{GGKS20}. It relies on
representing the partition function of the quantum chain at hand,
and more generally its statistical operator, by means of the partition
function of the underlying inhomogeneous vertex model with a special
choice of inhomogeneities (and boundary conditions). The method
is very powerful. It can as well be used for the calculation of
static \cite{GKS04a} and dynamic \cite{GKKKS17} correlation functions.
For a pedagogical introduction see \cite{Goehmann20}.

The calculation of correlation functions by means of Bethe Ansatz
has already a long history which could give rise to an encyclopedia
article on its own. Early success was connected with models that
have the same spectrum as free Fermions such as the XXZ chain
at $\D = 0$ or the Bose gas model at $c = \infty$. Leaving
these special cases aside, most of the relevant results that
were obtained so far pertain to the Bose gas or to the XXZ chain
at generic coupling or anisotropy. A most important ingredient of
these works is the Slavnov formula (\ref{slavnov}). The results
obtained in these works include multiple-integral representations
for ground state and finite temperature correlation functions,
e.g.\ \cite{KMT99b,KMST02a,GKS04a,GKS05}, factorized integrals
\cite{BGKS06} and results based on form-factor series for
the ground state \cite{Slavnov90,KMT99a,KKMST11b,KKMST12,BGKS21a}
and for finite temperatures \cite{KMS11a,KMS11b,DGK13a,GKKKS17},
to cite only a few of them. Powerful complementary methods have
been developed \cite{JiMi95,BJMST08a,JMS08} which are exact, but
rather use ideas from the representation theory of quantum
groups than the Bethe Ansatz.

\section{Omissions}
Bethe Ansatz has become a huge subject over the years. In this
essay we could only discuss a small part of it and necessarily had
to omit many interesting questions. Our exposition was mostly
historical. For this reason many recent works were not discussed.
Some subjects were left out at all, mostly because they do not overlap
with the expertise of the author. This includes, in particular,
many of the recent works on applications in quantum field theory
or in non-equilibrium problems which, moreover, have been
reviewed elsewhere. This also includes many interesting developments
on the side of the representation theory of quantum groups which are
sometimes closely, sometimes remotely related with the subject
we have discussed above.

\subsubsection*{Acknowledgment}
The author is grateful to M. D. Minin for a critical reading
of the manuscript and for his help with the figure.

\bibliographystyle{amsplain}
\bibliography{hub}

\end{document}